\documentclass[prb,twocolumn,showpacs,amsmath,amssymb,superscriptaddress]{revtex4-1}
\usepackage{graphicx}
\usepackage{dcolumn}
\usepackage{bm}
\usepackage{bbm}
\usepackage{ulem}
\usepackage{color}

\newcommand{\nc}{\newcommand}

\nc{\be}{\begin{equation}} \nc{\ee}{\end{equation}}
\nc{\bea}{\begin{eqnarray}} \nc{\eea}{\end{eqnarray}}
\nc{\bean}{\begin{eqnarray*}} \nc{\eean}{\end{eqnarray*}}

\begin{document}

\title{In-plane Magnetization Induced Quantum Anomalous Hall Effect}


\author{Xin Liu}
\affiliation{ Department of Physics, The Pennsylvania State University, University Park, Pennsylvania 16802-6300}
\author{Hsiu-Chuan Hsu}
\affiliation{ Department of Physics, The Pennsylvania State University, University Park, Pennsylvania 16802-6300}
\author {Chao-Xing Liu}\email{cxl56@psu.edu}
\affiliation{ Department of Physics, The Pennsylvania State University, University Park, Pennsylvania 16802-6300}

\date{\today}

\begin{abstract}
In a two-dimensional electron gas, the quantized Hall conductance can be induced by a strong magnetic field, known as the quantum Hall effect, and it can also result from the strong exchange coupling of magnetic ions, dubbed as the ``quantum anomalous Hall effect''. The quantum Hall effect requires the out-of-plane magnetic field, and similarly, it is commonly believed that the magnetization should be out-of-plane for the quantum anomalous Hall effect. In the present work, we find this condition is not necessary and predict that the quantum anomalous Hall effect can also be induced by the purely in-plane magnetization in two realistic systems, including Bi$_2$Te$_3$ thin film with magnetic doping and HgMnTe quantum wells with shear strains, when all the reflection symmetries are broken. An experimental setup is proposed to confirm this effect, the observation of which will pave the way to search for the quantum anomalous Hall effect in a wider range of materials.
\end{abstract}

\pacs{73.43.-f, 72.25.Dc, 75.50.Pp, 85.75.-d}
\maketitle

\newpage

For a two-dimensional (2D) electron gas in an out-of-plane magnetic field, a transverse voltage can be driven by the Lorentz force felt by electrons, known as the Hall effect, which was first discovered by E.H. Hall in 1879\cite{hall1879}. Later, Hall also observed the stronger transverse voltage in ferromagnetic conductors with an out-of-plane magnetization, dubbed as the anomalous Hall effect\cite{hall1881,nagaosa2010}, where the Hall effect is induced by the exchange coupling of magnetic ions. The Hall effect has its quantum version, the quantum Hall effect, in which the out-of-plane magnetic field is essential to form Laudau levels and obtain the quantized Hall conductance. In recent years, it is realized that the anomalous Hall effect also has its quantum version, dubbed as the quantum anomalous Hall (QAH) effect\cite{Haldane1988,onoda2003,Qi2006,Zhang2012}, where the quantized Hall conductance origins from the exchange coupling of magnetic ions instead of Landau levels of the out-of-plane magnetic fields. Recently, several realistic systems, including Mn doped HgTe quantum wells\cite{liu2009C}, magnetic impurities doped Bi$_2$Se$_3$ thin films\cite{yu2010}, GdBiTe$_3$ thin films\cite{zhang2011b}, {\it et al}\cite{wu2008,qiao2010}, have been proposed for the QAH effect and a large experimental effort have been made to persue the realization of this effect\cite{chang2011,zhang2012a,buhmann2002}.

The quantum Hall effect originates from the orbital effect of magnetic fields due to the Landau levels, which require the magnetic field along the z direction. Correspondingly, it is also commonly believed that the QAH effect requires z-direction magnetization. In-plane magnetic fields can not induce the orbital effect, however exchange couplings still exist for the in-plane magnetization. The breaking of time reversal (TR) symmetry, which is a necessary condition for the non-zero Hall conductance, occurs for any direction and any type of magnetization. Therefore, there is no constraint to limit the realization of the QAH effect with the in-plane magnetization. In the present work, we will show that the out-of-plane magnetization is not necessary and the QAH effect can also be induced by a purely in-plane magnetization. We propose two realistic systems to realize the in-plane magnetization induced QAH effect, which are accessible in the present experimental conditions.

We start from a general symmetry analysis of the necessary conditions for the appearance of non-zero Hall conductance. First, the Hall conductance must be zero in a TR invariant system, so a magnetic field or magnetization is required. Besides TR symmetry, the 2D point group (PG) symmetry gives an additional constraint for the Hall conductance, as first shown by Fang\cite{fang2012}. The 2D PGs consist of two families, the n-fold rotation symmetry $C_n$ and the n-fold dihedral symmetry $D_n$\cite{dresselhaus2008}. The dihedral group $D_n$ in 2D PGs is generated by the rotation $C_n$ and the reflection $M$. Here we emphasize that reflection $M$ in 2D PGs always corresponds to the reflection in three dimenional (3D) PGs with the reflection plane perpendicular to the 2D plane. The reflection in 2D PGs plays the role of inversion in 3D and distinguishes the pseudo-scalar (pseudo-vector) from the scalar (vector). The Hall conductance is zero if the 2D system has any reflection symmetry $M$. For example, let's consider a system with the reflection symmetry $M_x$ ($x\rightarrow -x, y\rightarrow y$) in the 2D plane, denoted as xy plane. For the Hall response $j_x=\sigma_{xy}E_y$, under $M_x$ the current $j_x$ changes its sign ($j_x\rightarrow -j_x$) while the electric field $E_y$ keeps its sign, so the Hall response equation is changed to $j_x=-\sigma_{xy}E_y$. If the system is invariant under $M_x$, the response equation should also be invariant under $M_x$, constraining the Hall conductance $\sigma_{xy}$ to be zero. Similar arguments can be applied to any 2D reflection symmetry. The out-of-plane magnetization is a pseudo-scalar in the 2D PGs, breaking any reflection symmetry $M$. In contrast, the in-plane magnetization, denoted as $\bm{m}$, is a pseudo-vector, and there is still a surviving reflection symmetry $M_m$ with the reflection plane perpendicular to $\bm{m}$, thus the in-plane magnetization by itself can not induce a non-zero Hall conductance and it is necessary to introduce other mechanisms to break the remaining reflection symmetry $M_m$. The symmetry analysis gives us a guidance to search for the non-zero Hall conductance with in-plane magnetization and in below, we will present two realistic systems, in which not only the non-zero Hall conductance, but also the quantum anomalous Hall effect can be realized with in-plane magnetization.

\textbf{Bi$_2$Te$_3$ thin film with magnetic doping}

Our first example is the Bi$_2$Te$_3$ thin film with magnetic doping, which possesses the QAH effect with an out-of-plane magnetization\cite{yu2010}. Here we will show that the QAH effect can also be induced by the in-plane magnetization in this system once we take into account the three-fold warping term\cite{fu2009c}. The low energy physics of a magnetically doped Bi$_2$Te$_3$ thin film is dominated by the two surface states on the top and bottom surfaces, with the Hamiltonian given by\cite{liu2010,yu2010}
\begin{eqnarray}\label{Ham-TI-1}
H&=&\left(\hbar v_F(\bm{\hat{z}}\times \bm{k})\cdot \bm{\sigma} +\frac{\lambda}{2}(k_+^3+k_-^3)\sigma_z\right)\tau_z+\bm{g}\cdot \bm{\sigma},
\end{eqnarray}
in the basis $|t\uparrow \rangle$, $|t\downarrow \rangle$, $|b\uparrow \rangle$ and $|b\downarrow \rangle$, where $t$ ($b$) is for the top (bottom)surface and $\uparrow$ ($\downarrow$) is for spin up (spin down). The Pauli matrices $\bm{\sigma}$ denote spin operators and $\tau_z=+1(-1)$ represent the surface states on the top (bottom) surface. We take the growth direction of the thin film as $\bm{\hat{z}}$ and the film plane as the xy plane. The first term is the kinetic term with the Fermi velocity $v_f$, the second term is the three-fold warping term with the parameter $\lambda$\cite{fu2009c} and the third term gives the spin splitting characterized by the parameters $\bm{g}=(g_x,g_y,g_z)$. The Zeeman type of spin splitting can originate from the direct Zeeman coupling between electron spin and magnetic fields, or from the exchange coupling between electron spin and magnetization of magnetic ions. The direction of $\bm{g}$ is along the in-plane magnetization $\bm{m}$. The hybridization between two surface states\cite{yu2010} is neglected here, which is not essential for our discussion below.
The QAH effect has been studied for this model when $\lambda=g_x=g_y=0$\cite{yu2010}, and here we focus on the case when both the in-plane magnetization and the warping term are non-zero.
The warping term breaks the full in-plane rotation symmetry down to the three-fold rotation ($C_3$) symmetry along $\bm{\hat{z}}$, which coincides with the symmetry of Bi$_2$Te$_3$ lattice in Fig. 1a.

\begin{figure}
\centering
\begin{tabular}{l}
\includegraphics[width=1.0\columnwidth]{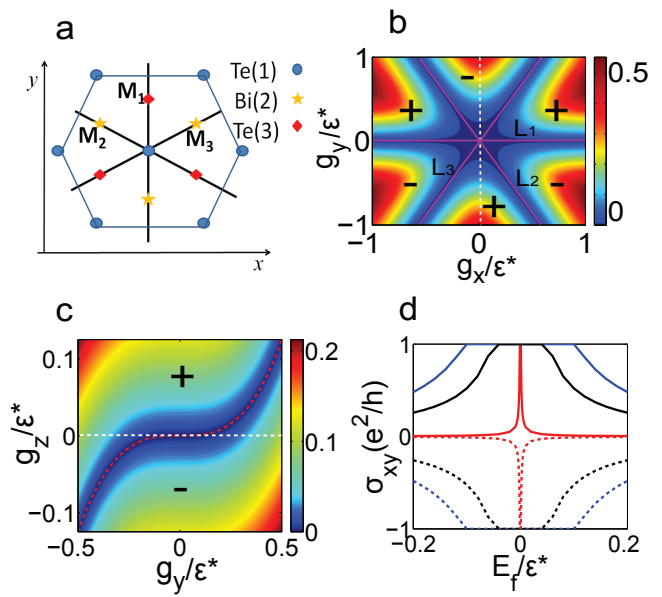}
\end{tabular}
\caption{\textbf{In-plane magnetization induced QAH effect in a Bi$_{\bm{2}}$Te$_{\bm{3}}$ film }. \textbf{(a)} Top view of Bi$_{\bm{2}}$Te$_{\bm{3}}$. The blue circles (red rhombuses) represent the Te atoms and the yellow stars represent the Bi atoms. The notion Te(1), Bi(2), Te(3) show the first three atomic layers away from the top surface. The three black lines (M$_1$, M$_2$ and M$_3$) indicate the three reflection planes. \textbf{(b)} The band gap as a function of $g_x$ and $g_y$. The $g_x-g_y$ plane is divided to six insulating regimes separated by the three metal lines ($L_{1,2,3}$) where the band gap is closed. The $+(-)$ in each area indicates the sign of the Hall conductance. Here the parameters $g_x$ and $g_y$ are rescaled with the energy unit $\varepsilon^*=v/\lambda=0.23$eV\cite{fu2009c}. \textbf{(c)} The band gap in $g_y-g_z$ plane. The red solid line indicates the band gap closing. The white dashed line corresponds to the white line in \textbf{(b)}. \textbf{(d)} The Hall conductance is plotted versus the chemical potential for different magnitudes of the in-plane magnetization. The blue, black, red solid lines correspond to the y direction magnetization $g_y=-0.5\varepsilon^*, -0.3\varepsilon^*, -0.1\varepsilon^*$, respectively, while the blue, black, red dashed lines correspond to $g_y=0.5\varepsilon^*, 0.3\varepsilon^*, 0.1\varepsilon^*$.  }
\label{sym-pha}
\end{figure}
The Bi$_2$Te$_3$ thin films have $D_3$ PG symmetry with three reflection operations $M_{1,2,3}$, which are related to each other by $C_3$ rotation, with the reflection planes indicated by three black lines in Fig. 1a.
One can easily check that the warping term in the Hamiltonian (\ref{Ham-TI-1}) preserves the x direction reflection $M_x=i\sigma_{x}$ ($M_x=M_1$) but breaks y direction reflection $M_y=i\sigma_y$, consistent with the lattice symmetry of Bi$_2$Te$_3$ in Fig. 1a.
According to the symmetry analysis, we requires the magnetization term breaking both the TR symmetry and $M_{1,2,3}$ in order to obtain a non-zero Hall conductance.

Since we are interested in the QAH regime, which requires an insulating state, it is instructive to check the band gap of the Hamiltonian (\ref{Ham-TI-1}), which shows a six-fold pattern as a function of the in-plane magnetization $g_x$ and $g_y$, as shown in Fig. 1b.
There are six insulating regimes with finite band gaps, which are separated by three gapless lines when the magnetization {\bf g} is along the direction indicated by the three lines $L_{1,2,3}$ in Fig. 1b.
This result coincides with the early calculation for a single surface\cite{Henk2012,Oroszlany2012}. Magnetization preserves the reflection symmetry $M_i$ ($i=1,2,3$) when its direction is along the line $L_i$. Therefore, we expect a zero Hall conductance for the magnetization along the gapless lines $L_{1,2,3}$. For the insulating regimes, the reflection symmetries $M_{1,2,3}$ are broken, so the Hall conductance can be non-zero. To determine the Hall conductance in the insulating regime, we consider how these insulating regimes are connected to the regimes with a finite out-of-plane magnetization $g_z$, where the quantized Hall conductance has been determined in the Ref. \cite{yu2010}. Fig. 1c
is plotted for the band gap as a function of $g_y$ and $g_z$, with the gap closing along the red dashed line. The white dashed line with $g_z=0$ in Fig. 1c
corresponds to the white dashed line in Fig. 1b
with $g_x=0$. From Fig. 1c, we find that the insulating regime with positive (negative) $g_y$ and $g_z=0$ is adiabatically connected to the regimes with negative (positive) $g_z$ and $g_y=0$. The system with a negative (positive) $g_z$ and $g_y=0$ should carry a Hall conductance $-\frac{e^2}{h}$ ($+\frac{e^2}{h}$), as shown in the Ref. \cite{yu2010}. Since the quantized Hall conductance can not vary for two adiabatically connecting insulating regimes, we expect the Hall conductance is $-\frac{e^2}{h}$ ($+\frac{e^2}{h}$) for the insulating regime with a positive (negative) $g_y$ and $g_{x,z}=0$. Due to the $C_3$ rotation symmetry, we can also determine the Hall conductance in other insulating regimes, as shown by the sign $\pm$ in Fig. 1b.
Moreover, we perform a direct calculation of the Hall conductance based on the Kubo formula \cite{thouless1982,Sinitsyn:2006_a}. As shown in Fig. 1d,
the Hall conductance is quantized when the Fermi energy lies within the band gap, confirming the above analysis. When the Fermi energy is above or below the band gap, the Hall conductance drops down, but is still non-zero. Based on the argument of adiabatically connection and the direct calculation of Hall conductance, we conclude that the quantized Hall conductance can be induced by the combination of the in-plane magnetization and the three-fold warping term in the present model.

The present argument based on the crystal symmetry also provides a way to distinguish the present mechanism from other possible mechanism for the anomalous Hall conductance, as shown by the experimental setup of the Hall measurement in Fig. 2a. \begin{figure}
\centering
\begin{tabular}{l}
\includegraphics[width=0.8\columnwidth]{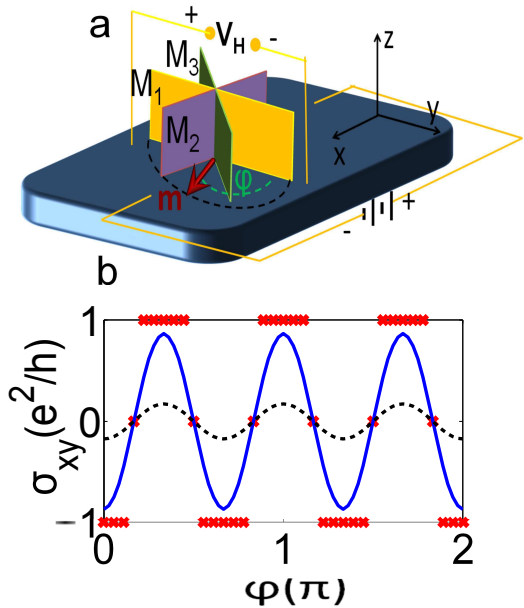}
\end{tabular}
\caption{\textbf{Experimental setup}. \textbf{(a)} The proposed experimental configuration to confirm the in-plane magnetization induced QAH effect. The arrow indicates the direction of magnetization. Three planes denote three reflection planes of the Bi$_2$Te$_3$ crystal. \textbf{(b)} The Hall conductance is plotted versus the angle of the in-plane magnetization at different Fermi energies. The red cross, black dashed and blue solid lines correspond to $E_f=0$, $-0.002\varepsilon^*$ and $0.02\varepsilon^*$ respectively. The Hall conductance oscillates between $\pm e^2/h$ with a $2\pi/3$ period no matter the chemical potential is in electron or hole band. Here we take $g_y=0.1\varepsilon^*$\cite{Henk2012}.  }
\label{sym-cond}
\end{figure}
By rotating the in-plane magnetization, the Hall conductance will switch between $\pm \frac{e^2}{h}$ for the insulating regimes, depending on the angle $\varphi$ between the magnetization and the crystal orientation, as shown in Fig. 2b.
Since there is no z direction magnetic field, the orbital effect can be safely excluded. The recent first principle calculation shows that the in-plane magnetization can open a gap around $0.25$meV on the surface of TIs\cite{Henk2012}. Therefore the QAH effect induced by the in-plane magnetization is expected below $3$K. For the metallic regimes, as shown by the blue solid and black dashed lines in Fig. 2b,
the Hall conductance is no longer quantized, but still oscillates between positive and negative values with $2\pi/3$ period of $\varphi$. This behavior is also different from that of the conventional anomalous Hall effect.

\textbf{Hg$_x$Mn$_{1-x}$Te quantum wells with shear strains}

The in-plane magnetization induced QAH effect is not limited to the concrete example of Bi$_2$Te$_3$ thin films, and instead, it can be generalized to other systems by engineering band structures. In the following, we will show how the shear strain can induce this effect in HgMnTe quantum wells and then discuss the general strategy to search for this effect.
The effective model for Mn doped HgTe quantum wells is described by the Bervenig-Hughes-Zhang (BHZ) Hamiltonian with an additional Zeeman type of coupling\cite{bernevig2006c,beugeling2012}
\begin{eqnarray}
	&&H=H_{BHZ}+H_{g}\\
	&&H_{BHZ}=\varepsilon(\bm{k})+\mathcal{M}(\bm{k})\tau_z+A(k_x\tau_x\sigma_z-k_y\tau_y)\label{eq:HamBHZ}\\
	&&H_{g}=\bm{g}_{1}\cdot\bm{\sigma}+\bm{g}_{2}\cdot\bm{\sigma}\tau_z
	\label{eq:HamZee}
\end{eqnarray}
in the basis $|E1+\rangle, |H1+\rangle, |E1-\rangle, |H1-\rangle$. Here the Pauli matrices $\bm{\sigma}$ are for spin and $\bm{\tau}$ are for the sub-bands of $E1$ and $H1$. The functions $\varepsilon(\bm{k})$, $\mathcal{M}(\bm{k})$, as well as the parameter $A$, are defined in Ref.\cite{bernevig2006c}. The vectors $\bm{g}_{1}=\frac{1}{2}(\bm{g}_{e}+\bm{g}_h)$ and $\bm{g}_{2}=\frac{1}{2}(\bm{g}_{e}-\bm{g}_h)$ with $\bm{g}_{e(h)}=(g_{e(h)x},g_{e(h)y},g_{e(h)z})$, which are treated as parameters in the following, describe the spin splitting for the $E1$ ($H1$) sub-band, and have the same direction as the in-plane magnetization $\bm{m}$.
The BHZ Hamiltonian with the magnetization along z direction ($g_{1(2)z}\neq 0$) has been studied in Ref. \cite{liu2008}, and the QAH phase is found when $g_{2z}$ is large enough. The BHZ Hamiltonian has the $D_{\infty}$ symmetry, so any plane perpendicular to the xy plane can serve as the reflection plane.
The in-plane magnetization $\bm{m}$ preserves the reflection symmetry $M_m$,
so the Hall conductance is zero for the BHZ model with the in-plane magnetization.

To obtain a non-zero Hall conductance, we need to break the remaining reflection symmetry, which can be achieved by introducing a new term due to the shear strains $\epsilon_{xz}$ and $\epsilon_{yz}$, written as
\begin{eqnarray}
	H_{str}=F\left[ \epsilon_{xz}(k_x\sigma_x+k_y\sigma_y)+\epsilon_{yz}(k_x\sigma_y -k_y\sigma_x) \right]\tau_x,
	\label{eq:Hamstrain}
\end{eqnarray}
with the parameter $F$. This form of the Hamiltonian can be derived from the six-band Kane model \cite{beugeling2012,novik2005}, as described in details in the supplementary on-line materials. $\epsilon_{xz}$ ($\epsilon_{yz}$) term breaks the x-direction reflection $M_x=i\sigma_{x}$ (the y-direction reflection $M_y=i\sigma_y\tau_z$) and preserves $M_y$ ($M_x$).
\begin{figure}
\centering
\begin{tabular}{l}
\includegraphics[width=1.0\columnwidth]{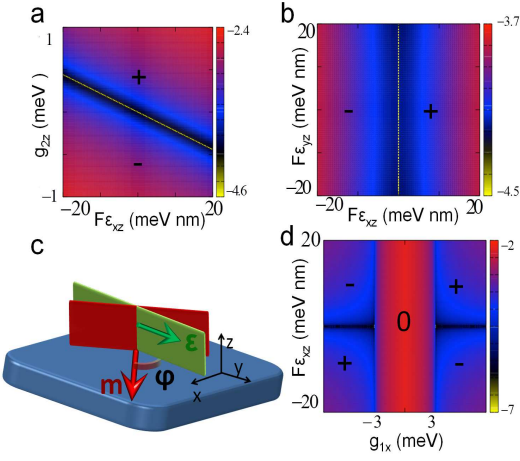}
\end{tabular}
\caption{\textbf{In-plane magnetization induced QAH effect in HgMnTe quantum wells}. \textbf{(a)} The band gap as a function of $g_{2z}$ and $\epsilon_{xz}$ with a finite $g_{1x}=8$ meV. \textbf{(b)} The band gap as a function of $\epsilon_{xz}$ and $\epsilon_{yz}$ with $g_{1x}=4$ meV, $g_{2x}=3$ meV. \textbf{(c)} Schematic plot of the experimental setup. The angle between the strain vector $\bm{\epsilon}$ (green arrow) and the magnetization vector $\bm{m}$ (red arrow) is denoted as $\varphi$. The green plane denotes the reflection plane preserved by the strain vector $\bm{\epsilon}$ while the red plane denotes the reflection plane preserved by the magnetization vector $\bm{m}$.  \textbf{(d)} The band gap as a function of $g_{1x}$ and $\epsilon_{xz}$ with $g_{2z}=0$. The other parameters are taken as $M=-3$ meV, $B=0.85$ eV$\cdot$nm$^2$, $D=0.67$ eV$\cdot$nm$^2$, $A=0.38$ eV$\cdot$nm, $g_{1(2)y}=g_{1z}=0$, and $\epsilon_{yz}=0$.     }
\label{HgMnTe}
\end{figure} Fig. 3a
shows the band gap for the Hamiltonian $H_{BHZ}+H_m+H_{str}$ as a function of $\epsilon_{xz}$ and $g_{2z}$ with a finite in-plane magnetization $g_{1x}$. When $\epsilon_{xz}=0$, the Hall conductance is $+\frac{e^2}{h}$ ($-\frac{e^2}{h}$) for positive (negative) $g_{2z}$, as obtained in Ref. \cite{liu2008}. The system is metallic for $g_{2z}=0$, separating the two QAH phases with opposite Hall conductances. With a finite $\epsilon_{xz}$, we find the gapless line derives away from the line of $g_{2z}=0$ and the regime with positive (negative) $\epsilon_{xz}$ and $g_{2z}=0$ is adiabatically connected to the regime with positive (negative) $g_{2z}$, which indicates that the Hall conductance for a postive (negative) $\epsilon_{xz}$ is $+\frac{e^2}{h}$ ($-\frac{e^2}{h}$), as shown in Fig. 3a. In Fig. 3b,
the band gap is plotted as a function of $\epsilon_{xz}$ and $\epsilon_{yz}$ with a finite $g_{1x}$,
and a gapless line along $\epsilon_{xz}=0$ separates two QAH phases with the Hall conductance $\pm\frac{e^2}{h}$. The Hall conductance vanishes along the gapless line, because both the shear strain $\epsilon_{yz}$ and the in-plane magentization $g_{1(2)x}$ preserves the reflection $M_y$. More generally, the Hall cnductance is always zero when two vectors, the shear strain $\bm{\epsilon}=(\epsilon_{xz},\epsilon_{yz})$ and the in-plane magnetization $\bm{m}$, are perpendicular to each other. We emphasize that the shear strain $\epsilon_{ij}$ is a tensor in 3D PGs, but we can treat $\bm{\epsilon}$ as a vector in 2D PGs. According to Fig. 3b, we can consider the experimental configuration for the magnetization and shear strain for the HgMnTe quantum wells, similar to that of Bi$_2$Te$_3$ thin films, as shown in Fig 3c.
When the angle $\varphi$ between the strain vector $\bm{\epsilon}$ and the magnetization vector $\bm{m}$ is rotated across $\frac{\pi}{2}$ or $\frac{3\pi}{2}$, the Hall conductance switches between $\pm\frac{e^2}{h}$ of the two insulating phases. In Fig 3d,
we verify the stability of the QAH phases for different values of $\bm{g}_1$ and the QAH phase always exists when the in-plane magnetization is large enough.

{\bf Discussion and conclusion}

From the above two examples, we find that the breaking of reflection symmetry is essential for the in-plane magnetization induced QAH effect. Generally, the pseudo-scalar, such as the out-of-plane magnetization, can break all the reflection symmetries in 2D PGs. Therefore, one should also construct a pseudo-scalar with the in-plane magnetization. For example, a pseudo-scalar, the dot product of a vector and a pseudo-vector $\bm{\epsilon}\cdot\bm{m}$, can be defined to characterize the Hall conductance in the HgMnTe quantum wells with shear strains. As shown in Fig. 3d, the sign of the Hall conductance is determined by the sign of the product of $\epsilon_{xz}$ and $g_{1x}$. We expect this strategy can also be applied to search for the QAH phase in other systems.

We propose the experiments with rotating in-plane magnetic fields, as shown in Fig. 2a
and Fig. 3c, to confirm the predicted effect. Since no out-of-plane magnetization is required, the orbital effect from Landau levels of magnetic fields can be excluded completely. Therefore, the proposed setups can provide a clear experimental signal to distinguish the orbital effect of magnetic fields from the exchange effect of magnetic ions. Our proposal is also feasible in experiments since Cr or Mn doped Bi$_2$Te$_3$ films or Mn doped HgTe quantum wells have already been realized \cite{chang2011,zhang2012a,buhmann2002}. Moreover, ferromagnetic materials are usually metallic, preventing the appearance of the QAH effect which requires insulating materials. The existing ferromagnetic insulators, such as EuO and GdN, have the in-plane magnetization for a thin film configuration\cite{Kasuya1997,steeneken2002,santos2008}.
Therefore, the in-plane magnetization induced QAH effect will pave the way to the new QAH materials with the hybrid structures made of ferromagnetic insulators.

\section*{ACKNOWLEDGMENTS}

We would like to thank Jainendra Jain, Laurens Molenkamp, Xiao-Liang Qi, Nitin Samarth, Yayu Wang, Qikun Xue and Shoucheng Zhang for useful discussions. X.L. acknowledges partial support by the DOE under Grant No. DE-SC0005042.

\newpage

\appendix
\section{}

In this appendix, we discuss the form of the effective four band model of HgMnTe quantum wells with shear strains based on the symmetry argument, as well as the microscopic derivation based on $k\cdot p$ theory.
%

The effective Hamiltonian of HgMnTe quantum wells with shear strains is constructed on the basis $|E1+\rangle$, $|H1+\rangle$, $|E1-\rangle$ and $|H1-\rangle$, including three terms $H_{BHZ}$, $H_m$ and $H_{str}$ (Eq. (2)-(5) in the main text of the article). The BHZ Hamiltonian $H_{BHZ}$ has been discussed in the early literature\cite{bernevig2006c,qi2011}. The most important feature of the BHZ Hamiltonian is the linear coupling between $|E1\pm\rangle$ and $|H1\pm\rangle$ due to the opposite parities between these two sub-bands.

The Hamiltonian $H_m$ is described by two vectors $\bm{g}_1=\frac{1}{2}(\bm{g}_e+\bm{g}_h)$ and $\bm{g}_2=\frac{1}{2}(\bm{g}_e-\bm{g}_h)$ (see the Eq. (4) in the main text), which can be related to the spin splitting of the electron and heavy-hole sub-bands. As described in the main text, $H_m$ has two origins: one is due to the direct Zeeman coupling to the external magnetic field, while the other orginates from the exchange coupling of magnetization. The direction of the vectors $\bm{g}_{1,2}$ or $\bm{g}_{e,h}$ are along the direction of the external magnetic fields, as well as the magnetization of the Mn magnetic moments. However, the dependence of $\bm{g}_{1,2}$ or $\bm{g}_{e,h}$ on the external magnetic field is quite complicted. Due to the quantum well configuration, there is also anisotropy between the spin splitting along the z-direction and that along the in-plane direction. For the z-direction, the form of spin splitting has already been obtained\cite{beugeling2012}, given by
\begin{eqnarray}
	g_{e(h)z}=\frac{\mu_B}{2}\tilde{g}_{e(h)z}B_z+\chi_{e(h)z}S_z
\end{eqnarray}
with the effective g factor $\tilde{g}_{e(h)z}$ and the exchange coupling strength $\chi_{e(h)}$ for the electron (heavy-hole) sub-bands, respectively. $S_z$ denotes the magnetization of Mn atoms $S_z=-S_0B_{5/2}\left( \frac{5g_{Mn}\mu_BB_z}{2k_B(T+T_0)} \right)$ where $S_0=5/2$, the effective g factor for Mn $g_{Mn}$, the Bhor magneton $\mu_B$, the characteristic temperature of the anti-ferromagnetic coupling $T_0$. $B_{5/2}(x)$ is the Brillioun function, given by $B_{5/2}(x)=\frac{6}{5}coth(\frac{6}{5}x)-\frac{1}{5}coth(\frac{1}{5}x)$. For the in-plane magnetic fields, the dependence of spin splitting on magnetic fiels is a little complicated. For the electron sub-band $|E1\pm\rangle$, the spin splitting is given by
\begin{eqnarray}
        g_{ex}+ig_{ey}=\frac{\mu_B}{2}\tilde{g}_{e\parallel}B_{+}+\chi_{e\parallel}S_+,
\end{eqnarray}
with the in-plane g factor $\tilde{g}_{e\parallel}$ and the in-plane exchange coupling strength $\chi_{e\parallel}$. Here $B_{\pm}=B_x\pm iB_y$ and $S_{\pm}=S_x\pm iS_y$ with $S_{x(y)}=-S_0 B_{5/2}\left(\frac{5g_{Mn}\mu_BB_{x(y)}}{2k_B(T+T_0)}\right)$. In contrast, the spin splitting for the heavy-hole sub-bands is given by
\begin{eqnarray}
        g_{hx}+ig_{hy}=\frac{\mu_B}{2}\tilde{g}_{h\parallel}B^3_{+}+\chi_{h\parallel}S^3_+.
\end{eqnarray}
The $B^3_+$ and $S^3_+$ dependence for the heavy-hole sub-bands is because the heavy-hole sub-bands carry the angular momentum $\pm\frac{3}{2}$.

For the shear strain term, the Hamiltonian (5) in the main text can be written explicitly as
\begin{widetext}
\begin{eqnarray}
H_{strain}=
F\left(
\begin{array}{cccc}
0 & 0 & 0 & k_-(\epsilon_{xz}-i\epsilon_{yz})\\
0 & 0 & k_-(\epsilon_{xz}-i\epsilon_{yz}) & 0 \\
0 & k_+(\epsilon_{xz}+i\epsilon_{yz}) & 0 & 0 \\
k_+(\epsilon_{xz}+i\epsilon_{yz}) & 0 & 0 & 0
\end{array}
\right)\label{app:Hstrain}
\end{eqnarray}
\end{widetext}
in the basis $|E1+\rangle$, $|H1+\rangle$, $|E1-\rangle$ and $|H1-\rangle$, with the shear strains $\epsilon_{xz}$ and $\epsilon_{yz}$ and the coefficient $F$. $\epsilon_{xz}$ and $\epsilon_{yz}$ have even parity and $\epsilon_{\pm}=\epsilon_{xz}\pm i\epsilon_{yz}$ carry the angular momentum $\pm 1$. Therefore, the term $k_\pm(\epsilon_{xz}\pm i\epsilon_{yz})$ carries the angular momentum $\pm 2$, corresponding to the change of the angular momentum between the electron sub-band $|E1+\rangle$ ($|E1-\rangle$) with the angular momentum $\frac{1}{2}$ ($-\frac{1}{2}$) and the heavy-hole sub-band $|H1-\rangle$ ($|H1+\rangle$) with the angular momentum $-\frac{3}{2}$ ($\frac{3}{2}$).

In order to confirm our discussion of the strain term, we may also consider the perturbation theory for the derivation of this term. We may start from the six band Kane model with the strain term, with the Hamiltonian given by
\begin{eqnarray}
	H=H_K+H_{str},
	\label{eq:H}
\end{eqnarray}
in the basis $|\Gamma^6,1/2\rangle$, $|\Gamma^6,-1/2\rangle$, $|\Gamma^8,3/2\rangle$, $|\Gamma^8,1/2\rangle$, $|\Gamma^8,-1/2\rangle$ and $|\Gamma^8,-3/2\rangle$, which we denote as $|1\rangle,|2\rangle,|3\rangle,|4\rangle,|5\rangle,|6\rangle$ for short in the following. In the conventional semiconductor, the $\Gamma^6$ bands have higher energy than the $\Gamma^8$ bands while in HgTe, the bands sequence is opposite.
The strain term can be described by the Bir-Pikus Hamiltonian with the substituion $k_ik_j$ in the Kane Hamiltonian of the strain tensor component $\epsilon_{ij}$. For the present model, we obtain
\begin{eqnarray}
	H_{str}=\left(
	\begin{array}{cccccc}
		T_\epsilon&0&0&0&0&0\\
		0&T_\epsilon&0&0&0&0\\
		0&0&U_\epsilon+V_\epsilon&S_\epsilon&R_\epsilon&0\\
		0&0&S^\dag_\epsilon&U_\epsilon-V_\epsilon&0&R_\epsilon\\
		0&0&R^\dag_\epsilon&0&U_\epsilon-V_\epsilon&-S_\epsilon\\
		0&0&0&R^\dag_\epsilon&-S^\dag_\epsilon&U_\epsilon+V_\epsilon
	\end{array}
	\right)
	\label{eq:Hstr}
\end{eqnarray}
with
\begin{eqnarray}
	&&T_\epsilon=C\mbox{tr}(\epsilon),\\
	&&U_\epsilon=a\mbox{tr}(\epsilon),\\
	&&V_\epsilon=\frac{1}{2}b(\epsilon_{xx}+\epsilon_{yy}-2\epsilon{zz}),\\
	&&S_\epsilon=-d(\epsilon_{xz}-i\epsilon_{yz}),\\
	&&R_\epsilon=-\frac{\sqrt{3}}{2}b(\epsilon_{xx}-\epsilon_{yy})+id\epsilon_{xy}.
	\label{eq:strainpara}
\end{eqnarray}
Here tr$(\epsilon)=\epsilon_{xx}+\epsilon_{yy}+\epsilon_{zz}$ gives the trace of the tensor $\epsilon$. Here we are interested in the influence of $\epsilon_{xz}$ and $\epsilon_{yz}$ and keep only $\epsilon_{xz}$ and$\epsilon_{yz}$ non-zero. From the above expressions, we find only $S_\epsilon$ terms remaining.

Next let's consider the projection of the Hamiltonian into the subspace of the quantum well sub-bands and derive the effective Hamiltonian for the low energy physics. We consider a quantum well structure frabricated along z direction and separate the Hamiltonian $H$ into two parts, $H=H_{k_\parallel=0}+H^{(1)}_{k_\parallel}$ with $k_\parallel=(k_x,k_y)$. $H_{k_\parallel=0}$ describes the Hamiltonian with vanishing $k_\parallel$ and is solved with numerical methods. The eigen wave functions of $H_{k_\parallel=0}$ is denoted as $|U_\xi(z)\rangle=\sum_\lambda f_{\xi,\lambda}(z)|\lambda\rangle$, where $|\lambda\rangle=|\Gamma^6,\pm\frac{1}{2}\rangle$ or $|\Gamma^8,\pm\frac{3}{2}(\pm\frac{1}{2})\rangle$. We use again 1 to 6 to denote $\lambda$ for short. $\xi$ denote different sub-bands and here we are interested in four sub-bands $|E_1,+\rangle$, $|H_1,+\rangle$, $|E_1,-\rangle$ and $|H_1,-\rangle$, denoted as $|A\rangle, |B\rangle,|C\rangle, |D\rangle$, with the forms given by
\begin{eqnarray}
	&&|A\rangle=f_{A,1}(z)|1\rangle+f_{A,4}(z)|4\rangle,\qquad |B\rangle=f_{B,3}|3\rangle\nonumber\\
	&&|C\rangle=f_{C,2}(z)|2\rangle+f_{C,5}(z)|5\rangle,\qquad |D\rangle=f_{D,6}|6\rangle.
\end{eqnarray}
The symmetry property of the functions $f_{\xi,\lambda}$ is discussed in the Ref. [\onlinecite{beugeling2012}]. We also consider other sub-bands, such as $|E_1,\pm\rangle$, $|LH,\pm\rangle$, $|HH_{2,3,\cdots},\pm\rangle$, in the second order perturbation calculation.

The effective Hamiltonian in the four bands (A,B,C,D sub-bands) can be obtained from the second order perturbation, which is illustrated in Ref. [\onlinecite{beugeling2012}] for the details. Here we are only interested in the terms related to the shear strains. For the first order term, the possible term is $\langle A|H_{str}|B\rangle=-(\epsilon_{xy}+i\epsilon_{yz})\langle f_{A,4}|d|f_{B,3}\rangle$, which vanishes because of the opposite party between $f_{A,4}$ and $f_{B,3}$ (see  Ref. [\onlinecite{beugeling2012}] for more details). Similar situation happens for $\langle C|H_{str}|D\rangle$. For the second order perturbation, the terms with the form $\langle A|H_{K}|LH,-\rangle\langle LH,-|H_{str}|D\rangle$ and $\langle B|H_{str}|LH,+\rangle\langle LH,-|H_{K}|C\rangle$ remains, yielding the Hamiltonian (\ref{app:Hstrain}), with the coefficient $F$ given by
\begin{widetext}
\begin{eqnarray}
	F=\sum_{\alpha=LH,E_2}\frac{1}{2\sqrt{6}}(\langle f_{A,1}|P|f_{\alpha^-,5}\rangle-\langle f_{A,4}|P|f_{\alpha^-,2}\rangle)\langle f_{\alpha^-,5}|d|f_{D,6}\rangle\left( \frac{1}{E_A-E_{\alpha^-}}+\frac{1}{E_D-E_{\alpha^-}} \right).
\end{eqnarray}
\end{widetext}
\newpage

%

%
%


\end{document}